\documentclass[11pt,a4paper]{article}
\pdfoutput=1

\usepackage{jheppub}
\usepackage{color}
\usepackage{graphicx}
\usepackage{wrapfig,enumerate,slashed}
\usepackage[utf8]{inputenc}

\usepackage{amsmath}
\usepackage{wasysym}
\usepackage{graphicx}
\usepackage{color}
\usepackage{comment}
\usepackage{hyperref}
\usepackage{epstopdf}
\usepackage{caption}
\usepackage{subcaption}

\newcommand{\be}{\begin{eqnarray}}
\newcommand{\ee}{\end{eqnarray}}

\newcommand{\bee}{\begin{eqnarray}}
\newcommand{\eee}{\end{eqnarray}}
\newcommand{\beeq}{\begin{equation}}
\newcommand{\eeeq}{\end{equation}}

\graphicspath{{Figures/}}

\begin{document}

\title{Electroweak oblique parameters as a probe of the trilinear Higgs boson self-interaction}

\author[a]{Graham D. Kribs,}
\author[b]{Andreas Maier,}
\author[c]{Heidi Rzehak,}
\author[b]{Michael Spannowsky,}
\author[b]{and Philip Waite}

\affiliation[a]{Department of Physics, University of Oregon, Eugene, OR, 97403 USA}
\affiliation[b]{Institute for Particle Physics Phenomenology, Department of Physics, Durham University, Durham, DH1 3LE, UK}
\affiliation[c]{CP3-Origins, University of Southern Denmark, Campusvej 55, DK-5230 Odense M, Denmark}

\emailAdd{kribs@uoregon.edu}
\emailAdd{andreas.maier@durham.ac.uk}
\emailAdd{rzehak@cp3.sdu.dk}
\emailAdd{michael.spannowsky@durham.ac.uk}
\emailAdd{p.a.waite@durham.ac.uk}

\abstract{We calculate the two-loop contributions from
a modified trilinear Higgs self-interaction,
$\kappa_\lambda \lambda_{\rm SM} v h^3$,
to the electroweak oblique parameters $S$ and $T$.
Using the current bounds on $S$ and $T$ from electroweak measurements,
we find the 95\% C.L. constraint on the modified trilinear coupling
to be $-14.0 \leq \kappa_\lambda \leq 17.4$.
The largest effects on $S$ and $T$ arise from two insertions
of the modified trilinear coupling that result in
$T/ S \simeq -3/2$; remarkably, this is nearly parallel
to the axis of the \emph{tightest} experimental constraint
in the $S$-$T$ plane.
No contributions to $S$ and $T$ arise from a modified Higgs
quartic coupling at two-loop order.
These calculations utilized a gauge-invariant parameterization
of the trilinear Higgs coupling in terms of higher-dimensional
operators $(H^\dagger H)^{n}$ with $n \ge 3$.
Interestingly, the bounds on $\kappa_\lambda$ that we obtain
are comparable to constraints from di-Higgs production
at the LHC as well as recent bounds from single Higgs production
at the LHC.}

\preprint{IPPP/17/15, CP3-Origins-2017-006 DNRF90}

\maketitle


\section{Introduction}
\label{sec:intro}

With the Higgs boson discovered \cite{Aad:2012tfa,Chatrchyan:2012xdj}, a major goal for current and future high-energy experiments is to provide precision measurements of Higgs couplings in order to thoroughly test the Standard Model and uncover any deviations.  A key ingredient to the Higgs mechanism \cite{Englert:1964et,Higgs:1964pj} is the shape and structure of the scalar potential, which, after spontaneous symmetry breaking, gives rise to trilinear and quartic Higgs self-interactions. The self-couplings of the Higgs boson are, at present, the least-constrained Higgs interactions of the Standard Model.  This motivates exploring a variety of techniques using a wide array of experimental data to constrain them.  In this paper, we evaluate how well electroweak precision data, expressed using the electroweak oblique parameters $S$ and $T$ \cite{Peskin:1990zt,Peskin:1991sw}, can constrain modifications of the trilinear Higgs self-interaction.

In the Standard Model (SM), the coefficients governing the shape of the scalar potential are determined by well-measured parameters in the broken phase -- the vacuum expectation value and the Higgs boson mass.  In order to study deviations from the SM, we consider a modified Higgs potential,
\begin{equation}
\label{higgsselfmod}
V_{\mathrm{mod}}(h) \supset \frac{m_h^2}{2}h^2 + \kappa_\lambda \lambda_{\mathrm{SM}} vh^3 + \kappa_4 \frac{\lambda_{\mathrm{SM}}}{4}h^4~,
\end{equation}
where only $m_h \simeq 125$~GeV\@ has been directly experimentally measured. In general, new physics that would result in modifications to the Higgs potential would also cause modifications to other couplings of the Standard Model.  In this paper, we consider \emph{only} the effects of modifying of the trilinear and quartic couplings \emph{in isolation} from the other Standard Model couplings. This is reasonable if we can formulate these modifications in a gauge-invariant way, and we can understand the impact of possible operator mixing through the renormalization group.

The formulation we use to implement the modified Higgs potential in Eq.~\eqref{higgsselfmod} is to add gauge-invariant higher-dimensional operators $\mathcal{O}_{2n} = -(H^{\dagger}H)^n/\Lambda^{2n-4}$ with cutoff scale $\Lambda$.  It will be convenient to write the coefficient as $1/\Lambda^{2n-4} \equiv \bar{c}_{2n} \lambda_{\mathrm{SM}}/v^{2n-4}$
(see e.g., Refs.~\cite{Giudice:2007fh,Contino:2013kra}).
In this formulation, the operators $\mathcal{O}_{2n}$ only affect the scalar potential, and moreover, $\mathcal{O}_6$ is known to not induce other dimension-6 operators under one-loop renormalization \cite{Elias-Miro:2013mua,Jenkins:2013zja}.  Hence, this modification  satisfies the requirements.  It is, however, also a potentially dangerous expansion since the Higgs potential receives corrections from higher-dimensional operators with, as we will see, $\Lambda \sim v$.  There are several possible ultraviolet completions of these higher-dimensional operators.  The simplest completion would involve new gauge singlets that interact with $H^\dagger H$ (but do not lead to singlet-Higgs boson mixing), such that integrating them out generates the tower of higher-dimensional operators. Other completions could lead to auxiliary modifications of other Higgs couplings; in this case, our analysis would be valid only if the other effects accidentally cancelled out leaving just the modified trilinear Higgs coupling.  In any case, our interest in this paper is to determine a model-independent bound on the trilinear Higgs coupling and leave the model-dependent interpretations to future work.

While the precise measurement of the trilinear Higgs self-coupling is highly challenging \cite{Baur:2002qd,Baur:2003gp,Dolan:2012rv,Baglio:2012np,Barr:2013tda,Dolan:2013rja,Papaefstathiou:2012qe,Goertz:2013kp,Maierhofer:2013sha,deLima:2014dta,Englert:2014uqa,Liu:2014rva,Goertz:2014qta,Azatov:2015oxa,Bishara:2016kjn}, first constraints have been obtained from direct searches for multi-Higgs boson final states at the LHC. Only recently have corrections from the modified trilinear couplings been considered in precision observables \cite{McCullough:2013rea,Degrassi:2017ucl}.  Furthermore, loop corrections to single Higgs production and associated Higgs production were used in Refs.~\cite{Degrassi:2016wml,Gorbahn:2016uoy,Bizon:2016wgr}.

In principle, we include the complete tower of operators $\mathcal{O}_{2n}$ given that the cutoff scale that we are considering is comparable to the vacuum expectation value of the Higgs boson.  In practice, we actually include just the effects of $\mathcal{O}_6$ on the explicit calculation of the electroweak oblique parameters $S$ and $T$.  This is not because the dimension-8 and higher order terms are unimportant, but instead one can show that the modified trilinear coupling captures the full effects of the tower of operators on $S$ and $T$ up to two-loop order.
Specifically, as we will see, no corrections to the quartic coupling enter our calculation of $S$ and $T$.   Consequently, to two-loop order, we can simply calculate corrections with $\mathcal{O}_6$ and reinterpret the correction,  without loss of generality, in terms of a modified trilinear coupling. A very clear discussion of this was also very recently presented in Ref.~\cite{Degrassi:2017ucl}.

There is another critical consequence of the observation that $S$ and $T$ do not depend on the modified quartic coupling to two loops.  Ordinarily, global questions of vacuum stability of the Higgs potential, such as whether the minimum is local or global, bounded from below, etc., may place severe constraints on the coefficients of a truncated theory, i.e., stopping at dimension-6 \cite{Datta:1996ni,Burgess:2001tj,Grojean:2004xa}.  Once at least dimension-8 terms are added, these concerns become parameter-dependent on the coefficients of the truncated tower.  This does not mean there are \emph{no} concerns with the stability of the potential -- only that these concerns require knowledge of the new physics beyond just the modified trilinear coupling.  If we stick to the ``high ground'' of model-independence, we can tacitly ignore Higgs potential stability issues.

In Sec.~\ref{sec:calculation} and~\ref{sec:oblique}, we outline our calculation of the effect that $\mathcal{O}_6$ has on the electroweak oblique parameters $S$ and $T$. We discuss the obtained limits, including a projection to future colliders, in Sec.~\ref{sec:results}, and present our conclusions in Sec.~\ref{sec:conclusions}.  We give the analytic expressions for $S$ and $T$ with the inclusion of the dimension-6 operator in Appendix~\ref{sec:resanalytic}.

\section{Higgs effective field theory and the modified Higgs potential}
\label{sec:calculation}

We begin by briefly reviewing the scalar potential in the SM
in order to define the SM couplings and the associated modifications.
The Higgs potential in terms of the Higgs doublet field $H$ is,
\begin{equation}
\label{smpotential}
V_{\mathrm{SM}}(H)=\mu_{\mathrm{SM}}^2H^{\dagger}H + \lambda_{\mathrm{SM}}(H^{\dagger}H)^2~.
\end{equation}
After electroweak symmetry breaking, the potential can be expanded around
the vacuum expectation value $v$ of the neutral component of the Higgs doublet,
$\mathrm{Re}[H^0] \equiv (h+v)/\sqrt{2}$.
The potential in terms of the physical Higgs field $h$
at the electroweak symmetry breaking minimum becomes,
\begin{equation}
\label{higgsself}
V_{\mathrm{SM}}(h) \supset \frac{m_h^2}{2}h^2 + \lambda_{\mathrm{SM}} vh^3 + \frac{\lambda_{\mathrm{SM}}}{4}h^4~,
\end{equation}
where $m_h^2 = -2\mu_{\mathrm{SM}}^2 = 2\lambda_{\mathrm{SM}} v^2$ and $v \simeq 246$~GeV.

The modified Higgs potential, Eq.~\eqref{higgsselfmod}, contains
the multiplicative factors $\kappa_\lambda$ and $\kappa_4$
that parameterizes the (potentially sizeable) corrections to the
trilinear and quartic couplings.
We implement the modified trilinear and quartic couplings using
higher-dimensional operators that only affect the Higgs potential,
\begin{eqnarray}
\label{higgsoperators}
\mathcal{L}_{\mathrm{EFT}} &=& - \sum_{n \geq 3} \frac{\bar{c}_{2n} \lambda_{\mathrm{SM}}}{v^{2n-4}} ( H^\dagger H )^{n}~,
\end{eqnarray}
where we have normalized the couplings with a factor of
$\lambda_{\mathrm{SM}} \equiv m_h^2/(2v^2)$.
The modified Higgs scalar potential becomes,
\begin{equation}
\label{dim6potential}
V(H) = \mu^2 H^{\dagger} H + \lambda (H^{\dagger}H)^2 + \sum_{n \geq 3} \frac{\bar{c}_{2n} \lambda_{\mathrm{SM}}}{v^{2n-4}} ( H^\dagger H )^{n}~,
\end{equation}
where now $\mu^2$ and $\lambda$ are in general different from
the SM values.

For now, consider extending the SM with just the additional dimension-6 operator $\mathcal{O}_6$. The minimization conditions are shifted, and so $\mu^2$ and $\lambda$ develop different relations in terms of the physical Higgs boson mass $m_h$ and vacuum expectation value $v$, which remain fixed to their experimental values. These relations are,
\begin{equation}
\label{dim6mulam}
\mu^2 = - \lambda_{\mathrm{SM}} v^2 \left( 1 - \frac{3}{4} \bar{c}_6 \right)~,
\qquad \lambda = \lambda_{\mathrm{SM}} \left( 1 - \frac{3}{2}\bar{c}_6 \right)~.
\end{equation}
Expanding the potential around the vacuum expectation value once again, the Higgs potential becomes Eq.~\eqref{higgsselfmod} with the identifications,
\begin{equation}
\label{kappac6relations}
\kappa_\lambda - 1 \; = \; \bar{c}_6~, \qquad
  \kappa_4 -1 \; = \; 6 \bar{c}_6~.
\end{equation}
At this stage, we have a gauge-invariant \emph{correlated} modification
of the trilinear and quartic Higgs self-couplings.
This can be generalized to two separate uncorrelated modifications
by including also the dimension-8 operator from Eq.~\eqref{higgsoperators}
with coefficient $\bar{c}_8$.
The modified trilinear and quartic Higgs self-couplings become,
\begin{equation}
 \kappa_\lambda -1 \; = \; \bar{c}_6 + 2 \bar{c}_8~, \qquad
 \kappa_4 - 1 \; =  \; 6 \bar{c}_6 + 16 \bar{c}_8~.
\end{equation}
 If we include even higher-dimensional operators $(H^\dagger H)^n$ with $n \le n_{\rm max}$, we again find two different linear combinations,
\begin{equation}
  \kappa_\lambda - 1 = \sum_{n= 3}^{n_{\rm max}} a_{2 n} \bar{c}_{2 n}~,
  \qquad \kappa_4 - 1 =   \sum_{n = 3}^{n_{\rm max}}  b_{2 n} \bar{c}_{2 n}~.
\end{equation}
The coefficients $a_{2 n}$ and $b_{2 n}$, where in general $a_{2 n} \neq b_{2 n}$, have to be evaluated for the chosen $n_{\rm max}$.
We will see that it is not necessary to include operators beyond the  additional dimension-6 operator $\mathcal{O}_6$ since the quartic
coupling, and hence $\kappa_4$, will be shown to not contribute to $S$ and $T$ at two loops. The result will therefore be expressed
in terms of $\bar{c}_6$, which will allow a direct translation in terms of the $\kappa_\lambda$ trilinear self-coupling modification.
Also, the higher-dimensional operators in Eq.~\eqref{higgsoperators}
generate even higher order Higgs boson interactions $\mathcal{O}(h^n)$ with $n \ge 5$, but since they do not contribute to the observables at the order to which we calculate, we do not need to consider them further.

\section{Electroweak oblique parameters}
\label{sec:oblique}

In the electroweak sector, the effect of new physics, if heavy, is expected to have its dominant contribution through the modification of gauge boson propagators via vacuum polarisation functions, or self-energies. These so-called oblique corrections can be parameterized in terms of the three Peskin-Takeuchi parameters, $S$, $T$ and $U$ \cite{Peskin:1990zt,Peskin:1991sw}. Since $U$ is only constrained by the $W$ boson mass and width, it is relatively insensitive to new physics, and so it is usually set to zero. $S$ and $T$ can therefore be used as a probe of the effects of new physics in the electroweak sector. They are defined by \cite{Degrassi:1993kn},
\begin{eqnarray}
\label{obliqueS}
S &=& \frac{4c^2s^2}{\alpha_e m_Z^2}\mathrm{Re}\left(\Pi_{ZZ}(m_Z^2)-\Pi_{ZZ}(0)-\frac{c^2-s^2}{cs}\left[\Pi_{Z\gamma}(m_Z^2)-\Pi_{Z\gamma}(0)\right]-\Pi_{\gamma\gamma}(m_Z^2)\right)~,\hspace{0.8cm} \\
\label{obliqueT}
T &=& \frac{1}{\alpha_e}\left(\frac{\Pi_{WW}(0)}{m_W^2}-\frac{c^2}{m_W^2}\left[\Pi_{ZZ}(0)+\frac{2s}{c}\Pi_{Z\gamma}(0)\right]\right)~.
\end{eqnarray}
In these equations, $\Pi_{AB}(p^2)$ represents the part of the self-energy proportional to the metric tensor $g^{\mu\nu}$ of the gauge boson $A$ propagating into the gauge boson $B$ with an external momentum $p$. $\alpha_e$ is the electromagnetic coupling constant, and we use the notation $s \equiv \sin\theta_W$ and $c \equiv \cos\theta_W$ where $\theta_W$ is the Weinberg angle. $S$ and $T$ are defined to arise solely due to the effects of new physics, and so when calculating these quantities, the SM contribution must be subtracted. The experimentally allowed values of the electroweak oblique parameters can be obtained by performing global fits to the electroweak precision observables and comparing the results to the SM prediction \cite{Baak:2014ora}.

Contributions to $S$ and $T$ involving the dimension-6 operator $\mathcal{O}_6$ first appear at the two-loop level. At this order in perturbation theory, self-energy diagrams containing both trilinear and quartic Higgs self-interactions appear, which due to their modifications from $\bar{c}_6$ outlined above, are manifest as non-zero corrections to $S$ and $T$. However, as we will see later, contributions from the quartic Higgs self-interaction exactly cancel in these observables. It is also important to note that at this order in perturbation theory, there are no vertex or box diagrams that depend on $\bar{c}_6$ involving light external fermions (i.e., light enough that their Yukawa couplings can be neglected). Since two-loop corrections to vertex or box diagrams involving both $\bar{c}_6$ and heavy external fermions do not enter the electroweak observables, the relevant two-loop $\bar{c}_6$ contributions to the self-energies must be separately gauge-invariant.

\subsection{Self-energy diagrams}

\begin{figure}[!tb]
\begin{subfigure}[b]{0.45\textwidth}
\includegraphics[width=\textwidth]{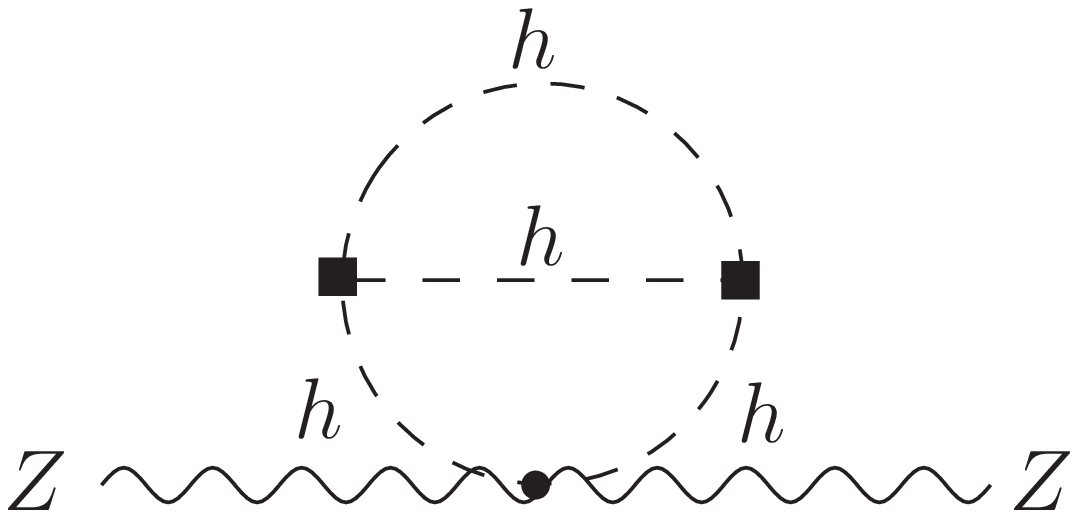}
\caption{}
\label{fig:ZZ_diagram}
\end{subfigure}
\begin{subfigure}[b]{0.45\textwidth}
\includegraphics[width=\textwidth]{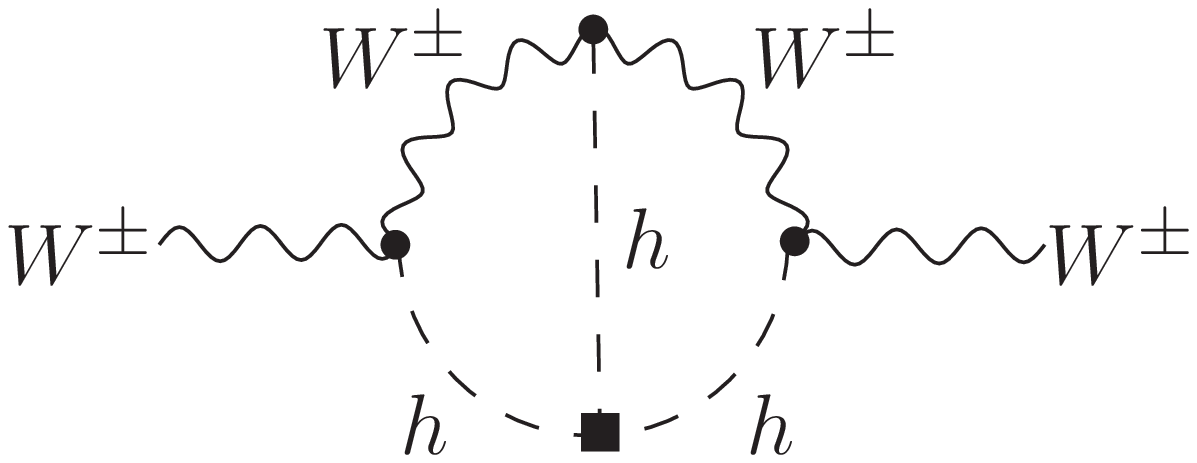}
\caption{}
\label{fig:WW_diagram}
\end{subfigure}
\begin{subfigure}[b]{0.45\textwidth}
\includegraphics[width=\textwidth]{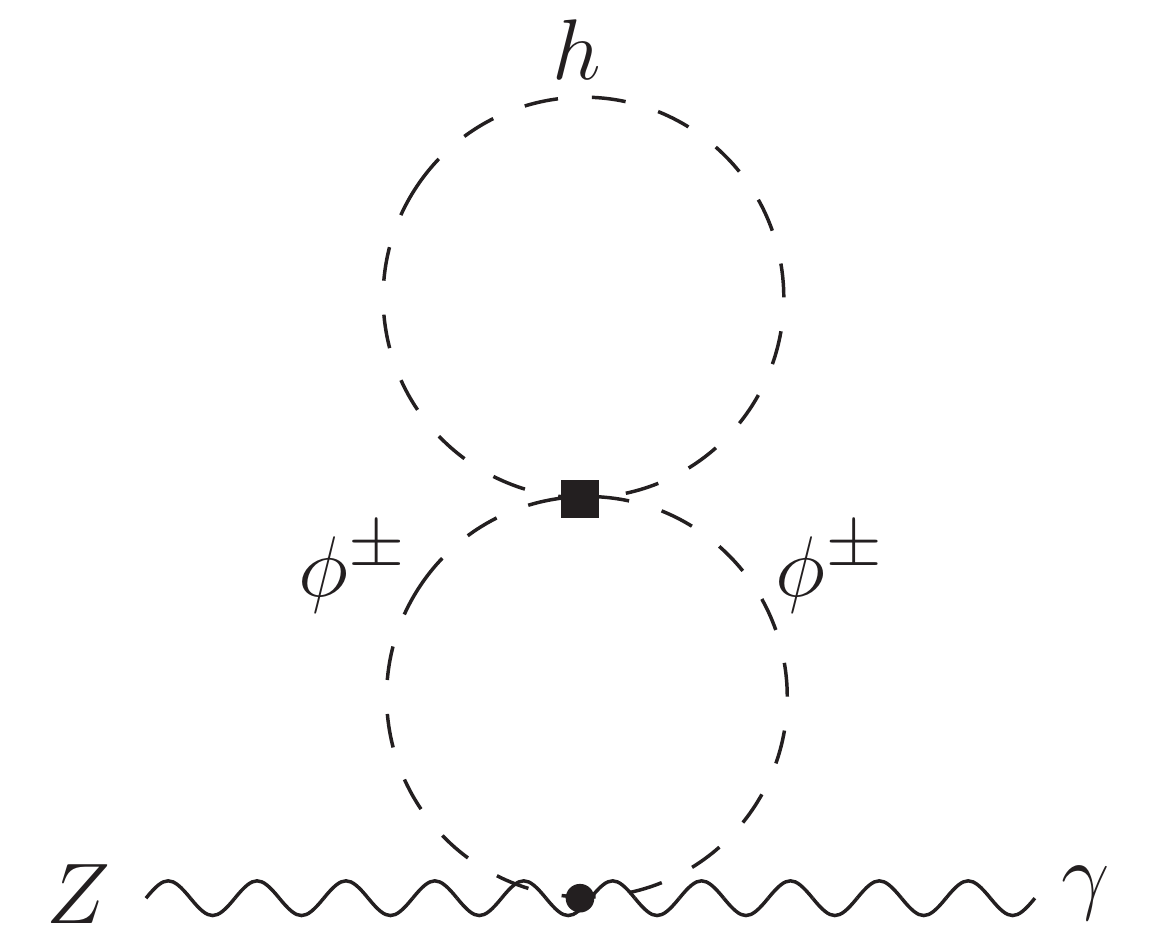}
\caption{}
\label{fig:Zgam_diagram}
\end{subfigure}
\hspace{1.5cm}
\begin{subfigure}[b]{0.45\textwidth}
\includegraphics[width=\textwidth]{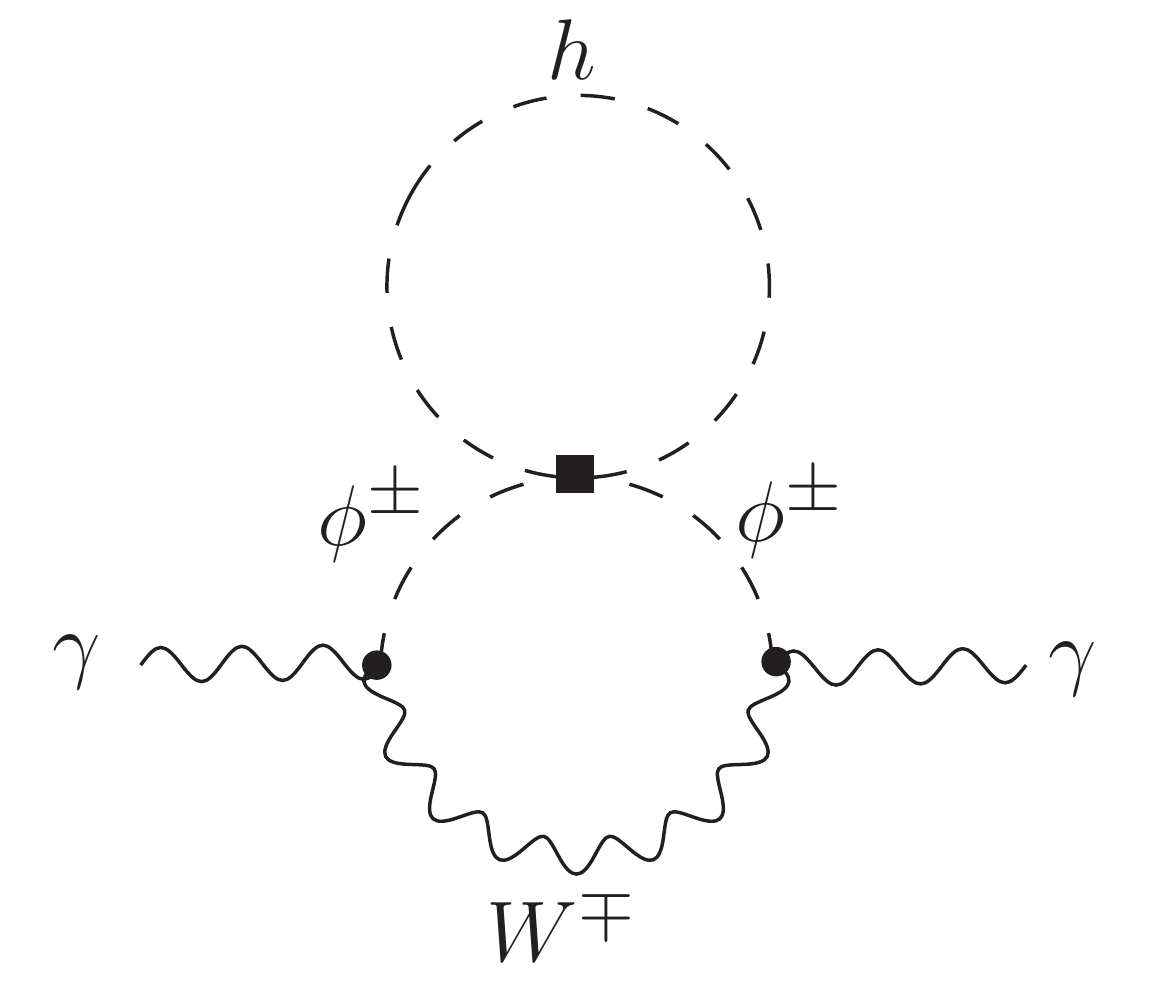}
\caption{}
\label{fig:gamgam_diagram}
\end{subfigure}
\caption{Example Feynman diagrams for the (a) $ZZ$, (b) $WW$, (c) $Z\gamma$ and (d) $\gamma\gamma$ two-loop self-energies. The square represents a vertex where there is a contribution from the dimension-6 operator.}
\label{fig:selfenergydiagrams}
\end{figure}

To evaluate the electroweak oblique parameters $S$ and $T$, all two-loop self-energy diagrams involving corrections from $\bar{c}_6$ need to be calculated. From the definitions of $S$ and $T$, all SM contributions are subtracted and so only terms proportional to $\bar{c}_6$ and $\bar{c}_6^2$ can remain. Working in the Feynman gauge, and discarding all two-loop diagrams that do not contain a contribution from $\bar{c}_6$, there are 26 diagrams for $ZZ$, 26 for $WW$, 5 for $Z\gamma$ and 5 for $\gamma\gamma$. An example Feynman diagram for each of the self-energies is shown in Fig.~\ref{fig:selfenergydiagrams}. From Eqs. \eqref{obliqueS} and \eqref{obliqueT}, it is apparent that the $ZZ$, $Z\gamma$ and $\gamma\gamma$ self-energies need to be evaluated at both zero and non-zero external momenta, whereas the $WW$ self-energies are only required with zero external momenta.

The two-loop self-energies can be reduced to linear combinations of a set of basis integrals using the reduction algorithm from O.V.~Tarasov \cite{Tarasov:1997kx}, based on integration by parts relations \cite{Chetyrkin:1981qh}. This reduction procedure is implemented in the \texttt{Mathematica} package \texttt{TARCER} \cite{Mertig:1998vk}, which is part of the program \texttt{FeynCalc} \cite{Mertig:1990an,Shtabovenko:2016sxi}. The amplitudes for the self-energy diagrams were generated using a model file in \texttt{FeynArts} \cite{Hahn:2000kx}, before using \texttt{TARCER} for the integral reduction. The reduction algorithm allows for the calculation of self-energies with non-zero external momenta and requires a total of eight basis integrals, but this reduces to a simplified set of two basis integrals when the external momenta are zero. A numerical implementation for the evaluation of all the basis integrals is given by the \texttt{TSIL} package \cite{Martin:2005qm}. The correspondence between the notations for the basis integrals in both \texttt{TARCER} and \texttt{TSIL} is given in the appendix of Ref.~\cite{Martin:2005qm}.

As a cross-check of our results, we have performed a second calculation of $S$ and $T$ based on an almost completely independent setup. After deriving the Feynman rules with the help of \texttt{FeynRules} \cite{Alloul:2013bka}, the self-energy diagrams were generated with \texttt{QGRAF} \cite{Nogueira:1991ex} and reduced to basis integrals using Laporta's algorithm \cite{Laporta:2001dd} as implemented in \texttt{FIRE} \cite{Smirnov:2014hma} and \texttt{Crusher} \cite{crusher}. Intermediate algebraic manipulations were performed with \texttt{FORM} \cite{Kuipers:2012rf}. Finally, the basis integrals were again evaluated numerically with \texttt{TSIL}. As a further check of our results, we used the \texttt{Mathematica} program \texttt{TwoCalc} \cite{Weiglein:1993hd} to verify the analytic expressions for the self-energy diagrams resulting from the \texttt{FeynArts} model file.

\subsection{Renormalization}

The leading order contribution to the electroweak oblique parameters from the Standard Model (and modifications to the renormalizable couplings) begins at one-loop.\footnote{We have assumed throughout the paper that the only higher-dimensional operators present are $\mathcal{O}_{2n}$, and in particular, the dimension-6 operators that give tree-level contributions to $S$ and $T$ are absent.} This means, for the calculation of these parameters at next-to-leading (two-loop) order, no actual two-loop counterterms are needed. However, all the tree-level parts entering into the one-loop leading order result, such as vertices and propagators, obtain a one-loop counterterm contribution in the next-to-leading order calculation of the oblique parameters. Since contributions of the $\mathcal{O}_6$ operator and the corresponding $\bar{c}_6$ parameter only enter at the two-loop level, no renormalization condition is needed for this parameter. All the other parameters are SM parameters, and we perform the renormalization procedure analogously to Ref.~\cite{Denner:1991kt}, which uses the on-shell scheme.

As already stated, we only take  $\bar{c}_6$-dependent corrections into account. Since the one-loop results for $S$ and $T$ are independent of $\bar{c}_6$, in order to obtain a  $\bar{c}_6$-dependent contribution at the two-loop level,  the one-loop counterterm insertions must depend on $\bar{c}_6$. In the counterterm vertices, the only $\bar{c}_6$-dependent contributions originate from the field renormalization constant of the Higgs boson, but these field renormalization constants cancel together with the  field renormalization constants from the counterterm insertions in the Higgs boson propagator. The only contributing counterterms are the Higgs mass and tadpole counterterms inserted into the Higgs boson and the Goldstone boson propagators.

\begin{figure}[!tb]
\begin{subfigure}[b]{0.3\textwidth}
\includegraphics[width=\textwidth]{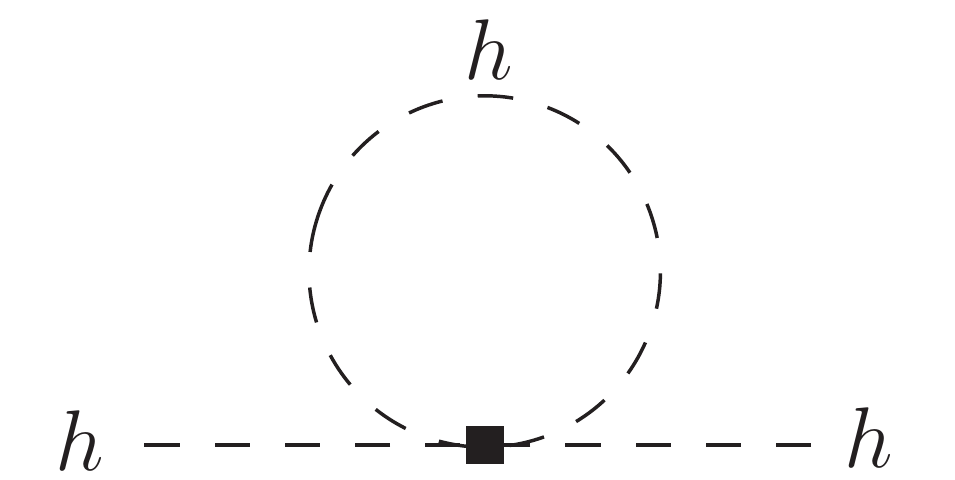}
\caption{}
\label{fig:quartic_higgs_mass_diagram}
\end{subfigure}
\begin{subfigure}[b]{0.3\textwidth}
\includegraphics[width=\textwidth]{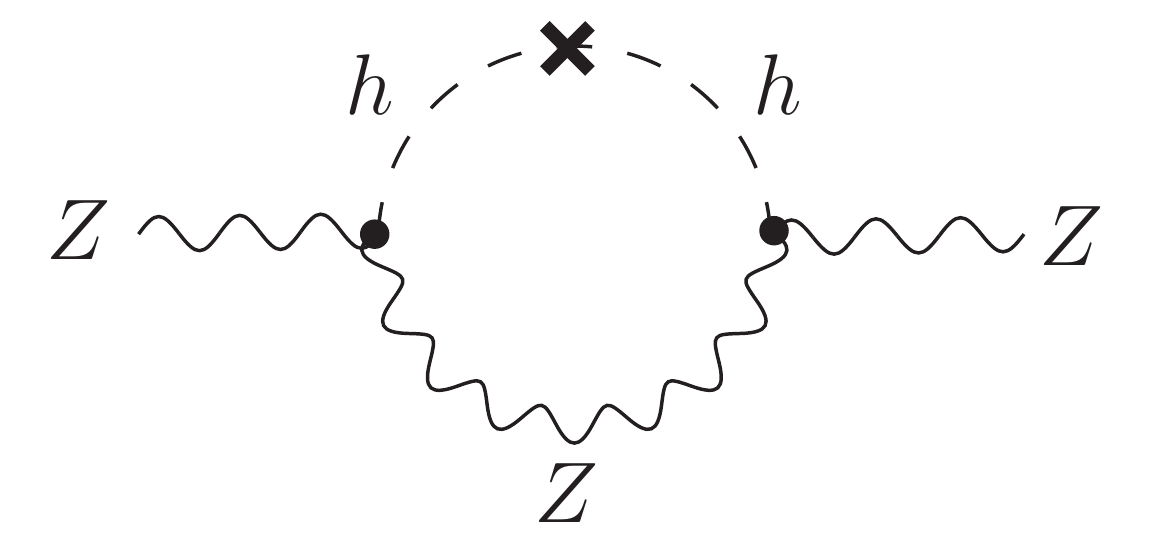}
\caption{}
\label{fig:counterterm_diagram}
\end{subfigure}
\begin{subfigure}[b]{0.3\textwidth}
\includegraphics[width=\textwidth]{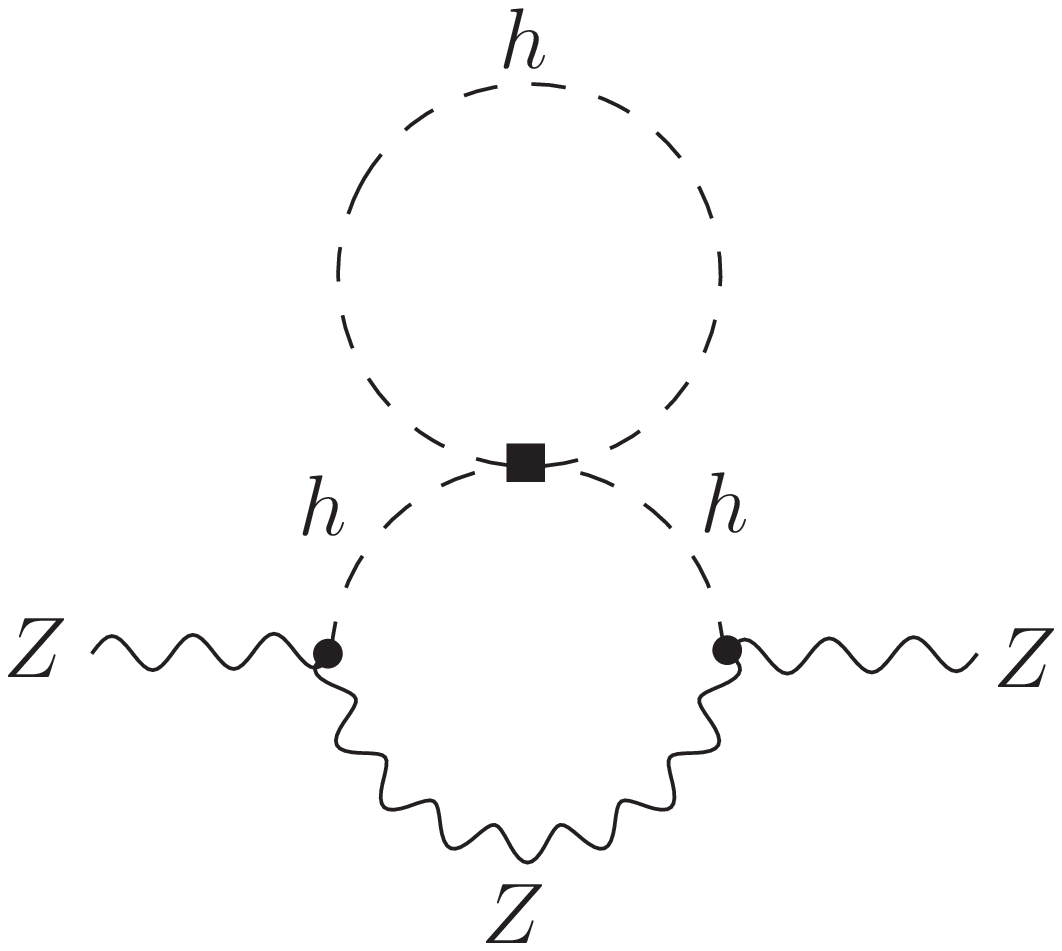}
\caption{}
\label{fig:quartic_higgs_diagram}
\end{subfigure}
\caption{Feynman diagrams demonstrating the cancellation of the quartic Higgs self-coupling. (a) shows the quartic contribution to the Higgs self-energy and (b) shows a counterterm insertion containing the quartic Higgs self-coupling which cancels with the contribution arising in (c). The square represents a vertex where there is a contribution from the dimension-6 operator, and the cross represents a counterterm insertion.}
\label{fig:quartic_cancellation}
\end{figure}

It should be noted that the counterterm insertion into the Higgs boson propagator contains a part that is proportional to the quartic Higgs self-coupling. It originates from the on-shell Higgs mass counterterm, $\delta m_h^2 = \Sigma_{hh}(m_h^2)$, and the corresponding contribution to the Higgs self-energy $\Sigma_{hh}$ shown in Fig.~\ref{fig:quartic_higgs_mass_diagram}. The correction due to Feynman diagrams with a counterterm insertion into the Higgs propagator, an example of which is shown in Fig.~\ref{fig:counterterm_diagram}, cancels the corresponding quartic Higgs self-couplings arising in the two-loop self-energy diagrams, such as in Fig.~\ref{fig:quartic_higgs_diagram}.

\section{Current and future limits from electroweak oblique parameters}\label{sec:results}

We have performed the calculation of the contribution from the dimension-6 operator $\mathcal{O}_6$ to the electroweak oblique parameters $S$ and $T$, and we find that after renormalization all ultraviolet divergences from the loop integrals cancel out, leaving non-zero and finite contributions to $S$ and $T$. Analytic expressions for the two-loop contributions to $S$ and $T$ from the $\bar{c}_6$ modification are given in Appendix~\ref{sec:resanalytic}. For the numerical analysis, we take as input parameters \cite{Olive:2016xmw}:
\begin{align}
m_W &= 80.385~\mathrm{GeV}~, & m_Z &= 91.1876~\mathrm{GeV}~, \nonumber\\
m_h &= 125~\mathrm{GeV}~, & G_F &= (1.16637870\times 10^{-5})~\mathrm{GeV}^{-2}~.
\end{align}
The $W$ and $Z$ boson masses are the pole masses, and the electroweak scheme is specified by the tree-level relations between the parameters \cite{Sirlin:1980nh}. We find that the contribution of $\bar{c}_6$ to $S$ and $T$ is,
\begin{eqnarray}
S &=& - 0.000138~\bar{c}_6^2 + 0.000180~\bar{c}_6~, \nonumber\\
T &=& \phantom{-}0.000206~\bar{c}_6^2 - 0.000324~\bar{c}_6~.
\end{eqnarray}
As there are no contributions from the quartic Higgs self-coupling, we can use the relation between $\bar{c}_6$ and $\kappa_\lambda$ in Eq.~\eqref{kappac6relations} to write this result as,
\begin{eqnarray}
S &=& -0.000138~(\kappa_\lambda^2 - 1) + 0.000456~(\kappa_\lambda - 1)~,
\nonumber\\
T &=& \phantom{-}0.000206~(\kappa_\lambda^2 - 1) - 0.000736~(\kappa_\lambda - 1)~.
\end{eqnarray}
The distinction between the contribution from two insertions of a modified Higgs self-coupling and a single insertion is made explicit here, since a term proportional to $(\kappa_\lambda^2 - 1)$ is exactly the contribution we get from two insertions.

\begin{figure}[!tb]
\centering
\includegraphics[width=0.7\textwidth]{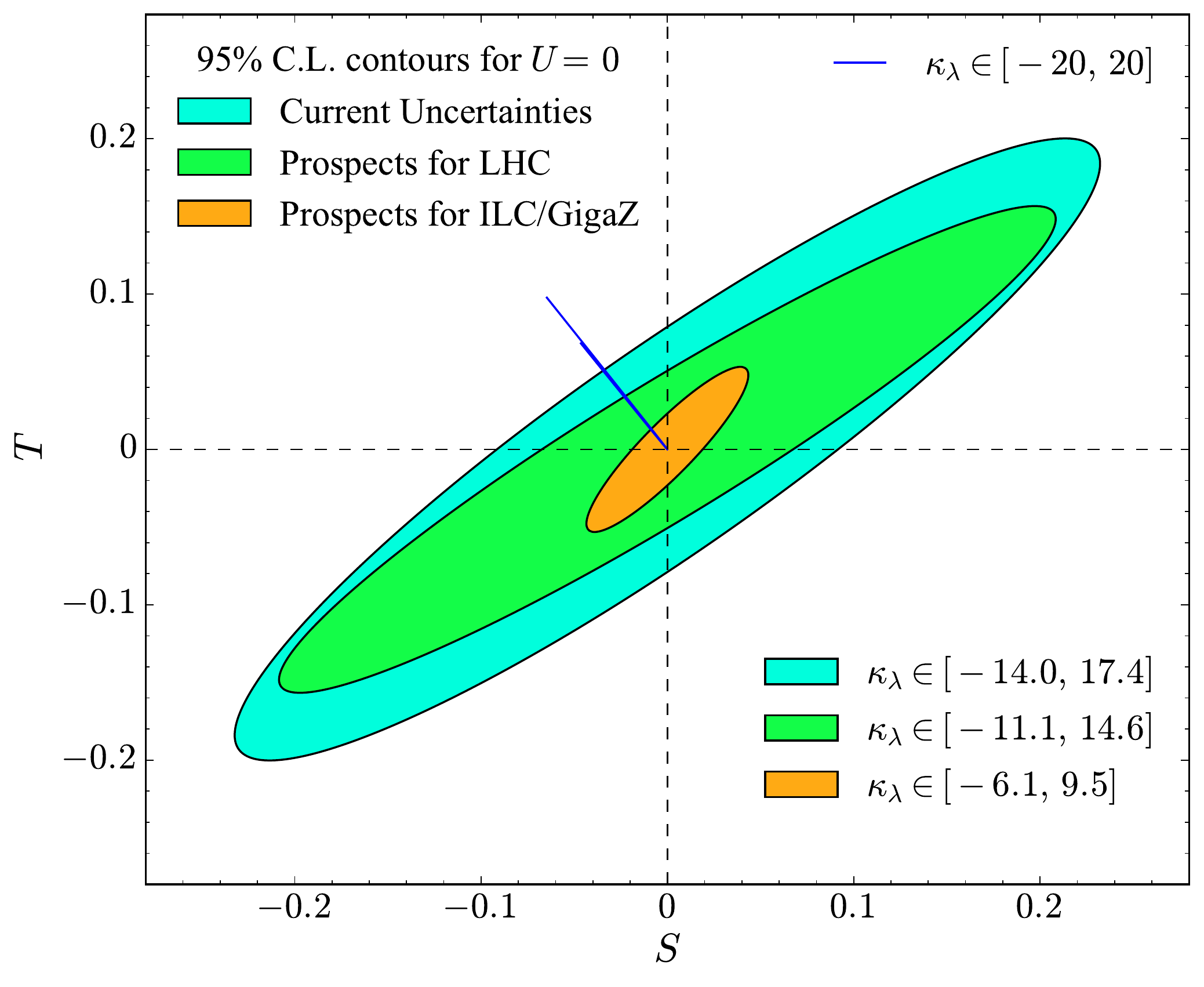}
\caption{Current limits and projected sensitivities of $\kappa_\lambda$ from the electroweak oblique parameters $S$ and $T$. The light blue area in the $S$-$T$ plane corresponds to the 95\% C.L. region based on measurements at LEP and the LHC. The green and orange areas correspond to projected LHC and ILC/GigaZ sensitivities respectively.  The longer (shorter) thin blue lines show the shift in $S$ and $T$ as $\kappa_\lambda$ extends up to $-20$ ($+20$). The intersection of these lines with the current limits and projected sensitivities gives the ranges of $\kappa_\lambda$ as shown in the figure.}
\label{fig:ST_full_kappa}
\end{figure}

The path of the $\kappa_\lambda$ contribution in the $S$-$T$ plane is shown in Fig.~\ref{fig:ST_full_kappa}. The light blue ellipse shows the current 95\% C.L. bound on the $S$ and $T$ parameters, as obtained by The Gfitter Group \cite{Baak:2014ora}. Also shown in the plot are possible future bounds on these parameters. The ellipses are constructed for $U=0$ and are centred on $(0,0)$. From the intersection points of the path of $\kappa_\lambda$ in the $S$-$T$ plane with the current ellipse, we estimate for the 95\% C.L. a bound of:
\begin{equation}
 -14.0 \leq \kappa_\lambda \leq 17.4~.
 \label{eq:res}
\end{equation}
Similar bounds have been derived using the observables $m_W$ and $\sin\theta_W$ instead of $S$ and $T$ \cite{Degrassi:2017ucl}.
The limits of Eq.~\eqref{eq:res} can be compared to existing bounds from searches for di-Higgs final states and Higgs coupling measurements. Direct searches constrain $\kappa_\lambda$ to $-14.5 \leq \kappa_\lambda \leq 19.1$ \cite{Aad:2015xja,Gorbahn:2016uoy} and $-8.4 \leq \kappa_\lambda \leq 13.4$ \cite{ATLAS-CONF-2016-049, Bizon:2016wgr} using Run I and Run II data respectively. In addition, Higgs coupling measurements performed in single Higgs production result in the combined bound of $-9.4 \leq \kappa_\lambda \leq 17.0$ \cite{Degrassi:2016wml}. While current limits from single Higgs production are stronger than bounds derived from electroweak precision measurements, they provide complementary information and can be used to extract a combined limit.

\section{Conclusions}
\label{sec:conclusions}

Detailed knowledge of the self-interactions of the Higgs boson is of crucial importance to improve our understanding of the underlying mechanism of electroweak symmetry breaking and the nature of the Higgs boson itself. Only very recently have investigations of constraints on the trilinear  self-interaction from (di-)Higgs production at the LHC begun to appear. However, in the absence of a signal in di-Higgs production (and thus a  determination of the Higgs self-coupling), alternative ways of studying Higgs self-interactions can help to shed light on the dynamics of the scalar interactions of the Higgs boson. For example, loop-induced single Higgs production has recently been investigated and found to provide comparable limits to those from di-Higgs cross section measurements.

In this study, we have focused on the effect of Higgs self-interactions on the electroweak oblique parameters $S$ and $T$ in order to set limits on a modified trilinear self-coupling. Since the self-energies needed for $S$ and $T$ do not involve external Higgs bosons, the effects of a modified trilinear self-coupling appear only at the two-loop level and above. We found that at this order the quartic Higgs self-coupling has no effect, enabling us to set model-independent limits on $\kappa_\lambda$ from its effects on $S$ and $T$ using a gauge-invariant effective field theory approach.

Our estimate for the current 95\% C.L. bound on $\kappa_\lambda$ is comparable to bounds derived from single Higgs processes. As the two approaches are orthogonal in nature, with independent uncertainties, they can be used to check the self-consistency of the bounds and, in combination
(see e.g., Ref.~\cite{Degrassi:2017ucl}), set better limits on the
trilinear Higgs self-coupling.

Reinterpreting limits on $\kappa_\lambda$ as bounds on the scale of the higher-dimensional operators $(H^\dagger H)^n/\Lambda^{2n - 4}$ implies a lower bound on the cutoff scale of order $\Lambda \gtrsim (v/\sqrt{2}) \times \sqrt{15.5/\bar{c}_6}$. Should evidence for such a large deviation in the Higgs trilinear self-coupling appear, this clearly implies the scale of the new physics must be very close to the scale of electroweak breaking. The simplest models of new physics would involve singlets that couple only to $(H^\dagger H)$ but without mixing with the Higgs boson.  We leave for future work the investigation of such models, and whether they could permit large deviations in the trilinear self-coupling without having appeared in any other collider search. It is tempting to also consider the implications on the electroweak phase transition. The presence of the dimension-6 operator with $\bar{c}_6 \lesssim 2$ has been known for some time to suggest the transition becomes first-order
\cite{Grojean:2004xa,Noble:2007kk}. Larger values of $\bar{c}_6$ run into trouble with the global properties of the Higgs potential (global vs local minimum), but obviously once $\bar{c}_6$ is large enough to suggest $\Lambda$ is near the electroweak scale, it no longer makes sense to truncate to dimension-6.  Should evidence for large deviations in the trilinear self-coupling be observed, the electroweak phase transition is undoubtedly drastically modified. If new physics causing deviations in the trilinear self-coupling at the level that could be probed from future electroweak precision tests existed so close to the electroweak scale, it seems unavoidable that the full theory realizing the effects of the effective operators is needed to fully understand and characterize the electroweak phase transition.

\emph{Note added:}  As this paper was being completed,
Ref.~\cite{Degrassi:2017ucl}, which also considered
electroweak precision bounds on the trilinear Higgs self-coupling, appeared.
Their approach was to calculate the two-loop contributions to
$m_W$ and $\sin^2\theta^{\rm lep}_{\rm eff}$, and the bounds
they obtained (at 95\% C.L.) can be read off from their Fig.~4,
roughly $-14 \leq \kappa_\lambda \leq 17$, fully consistent
with our results.

\vskip 1 \baselineskip

\acknowledgments
The authors thank P. Marquard for providing the program \texttt{Crusher} for integral reduction. GDK is supported in part by the U.S. Department of Energy under Grant No. DE-SC0011640. AM is supported by a European Union COFUND/Durham Junior Research Fellowship under EU grant agreement number 267209. HR's work is partially funded by the Danish National Research Foundation, grant number DNRF90. MS is supported in part by the European Commission through the ``HiggsTools'' Initial Training Network PITN-GA-2012-316704.


\appendix

\section{Analytic results}
\label{sec:resanalytic}
\allowdisplaybreaks
In the following, we present the analytic results for the $\bar{c}_6$ contributions to $S$ and $T$.
The notation for the basis integrals closely
follows Ref.~\cite{Martin:2005qm}. For the self-energy diagrams $B$, $S$, $T$, $U$,
and $M$, the first argument is the square of the external momentum,
\begin{align}
  \label{eq:S_analytic}
S = {}&\frac{\alpha_e\*\bar{c}_6}{1024\*\pi^2\*s^2\*m_W^2\*m_Z^4\*(m_h^2 - 4\*m_Z^2)\*(m_h^2 - m_Z^2)^2}\*\bigg\{\notag\\
&\ +
36\*(2+\bar{c}_6)\*m_h^2\*(m_h^2-m_Z^2)\*B(m_h^2,m_h^2,m_h^2)\*\Big(-m_Z^2\*(m_h^6-3\*m_h^4\*m_Z^2+4\*m_h^2\*m_Z^4+16\*m_Z^6)\notag\\
&\quad+2\*(m_h^2-2\*m_Z^2)^3\*(m_h^2-m_Z^2)\*B(m_Z^2,m_h^2,m_Z^2)\Big)\notag\\
&\ +8\*m_h^2\*A(m_Z^2)\*\Big(-4\*(m_h^2-4\*m_Z^2)\*(m_h^2-2\*m_Z^2)\*(m_h^2-m_Z^2)^2\*B(m_Z^2,m_h^2,m_Z^2)\notag\\
&\quad-(m_h^2-m_Z^2)\*\big[(10+3\*\bar{c}_6)\*m_h^6-3\*(18+5\*\bar{c}_6)\*m_h^4\*m_Z^2\notag\\
&\qquad+48\*(3+\bar{c}_6)\*m_h^2\*m_Z^4-4\*(34+9\*\bar{c}_6)\*m_Z^6\big]\notag\\
&\quad-9\*(2+\bar{c}_6)\*(m_h^8-6\*m_h^6\*m_Z^2+14\*m_h^4\*m_Z^4-8\*m_h^2\*m_Z^6+8\*m_Z^8)\*B(m_h^2,m_h^2,m_h^2)\Big)\notag\\
&\ +8\*A(m_h^2)\*\Big(-2\*m_h^2\*(2\*m_h^6-9\*m_h^4\*m_Z^2+16\*m_Z^6)\*A(m_Z^2)\notag\\
&\quad-m_h^2\*(m_h^2-m_Z^2)\*\big[(14+3\*\bar{c}_6)\*m_h^6-6\*(10+\bar{c}_6)\*m_h^4\*m_Z^2\notag\\
&\qquad+12\*(7+\bar{c}_6)\*m_h^2\*m_Z^4+8\*(20+9\*\bar{c}_6)\*m_Z^6\big]\notag\\
&\quad+2\*(m_h^2-4\*m_Z^2)\*(m_h^2-m_Z^2)^2\*\big[(4+\bar{c}_6)\*m_h^4-4\*(3+\bar{c}_6)\*m_h^2\*m_Z^2\notag\\
&\qquad+12\*(2+\bar{c}_6)\*m_Z^4\big]\*B(m_Z^2,m_h^2,m_Z^2)\notag\\
&\quad+9\*(2+\bar{c}_6)\*m_h^2\*(m_h^8-7\*m_h^6\*m_Z^2+19\*m_h^4\*m_Z^4-24\*m_h^2\*m_Z^6+20\*m_Z^8)\*B(m_h^2,m_h^2,m_h^2)\Big)\notag\\
&\ +m_h^2\*\bigg(-8\*\big[(12+7\*\bar{c}_6)\*m_h^8-9\*(9+5\*\bar{c}_6)\*m_h^6\*m_Z^2+99\*(2+\bar{c}_6)\*m_h^4\*m_Z^4\notag\\
&\qquad-8\*(15+8\*\bar{c}_6)\*m_h^2\*m_Z^6+12\*(6+7\*\bar{c}_6)\*m_Z^8\big]\*I(m_h^2,m_h^2,m_h^2)\notag\\
&\quad+(m_h^2-4\*m_Z^2)\*\Big[8\*(4\*m_h^4-5\*m_h^2\*m_Z^2-2\*m_Z^4)\*A(m_Z^2)^2\notag\\
&\qquad+24\*m_Z^2\*\Big((m_h^2-2\*m_Z^2)\*A(m_h^2)^2+m_Z^2\*\big[20\*m_Z^2-(20+9\*\bar{c}_6)\*m_h^2\big]\*I(m_h^2,m_h^2,m_Z^2)\Big)\notag\\
&\qquad-8\*(m_h^2+2\*m_Z^2)\*(2\*m_h^4-9\*m_h^2\*m_Z^2+16\*m_Z^4)\*I(m_h^2,m_Z^2,m_Z^2)\Big]\notag\\
&\quad+(m_h^2-m_Z^2)\*\Big\{128\*m_h^8+32\*\bar{c}_6\*m_h^8-554\*m_h^6\*m_Z^2-99\*\bar{c}_6\*m_h^6\*m_Z^2+986\*m_h^4\*m_Z^4\notag\\
&\qquad+279\*\bar{c}_6\*m_h^4\*m_Z^4+432\*m_h^2\*m_Z^6+268\*\bar{c}_6\*m_h^2\*m_Z^6+304\*m_Z^8+168\*\bar{c}_6\*m_Z^8\notag\\
&\qquad+8\*(m_h^2-m_Z^2)\*\Big[\notag\\
&\qquad\ +\big[(22+9\*\bar{c}_6)\*m_h^4-12\*(8+3\*\bar{c}_6)\*m_h^2\*m_Z^2+8\*(22+9\*\bar{c}_6)\*m_Z^4\big]\*S(m_Z^2,m_h^2,m_h^2,m_Z^2)\notag\\
&\qquad\ +8\*(m_h^2-m_Z^2)\*\big[(4+\bar{c}_6)\*(m_h^4-4\*m_h^2\*m_Z^2)+12\*(2+\bar{c}_6)\*m_Z^4\big]\*T(m_Z^2,m_h^2,m_h^2,m_Z^2)\notag\\
&\qquad\ +(m_h^2-4\*m_Z^2)\*\Big(2\*(m_h^6-12\*m_h^2\*m_Z^4+24\*m_Z^6)\*M(m_Z^2,m_h^2,m_h^2,m_Z^2,m_Z^2,m_h^2)\notag\\
&\qquad\quad+\big[(2+\bar{c}_6)\*m_h^4-4\*(2+\bar{c}_6)\*m_h^2\*m_Z^2+4\*(10+3\*\bar{c}_6)\*m_Z^4\big]\*B(m_Z^2,m_h^2,m_Z^2)\notag\\
&\qquad\quad-4\*m_Z^2\*(m_h^2-2\*m_Z^2)\*\big[B(m_Z^2,m_h^2,m_Z^2)^2+2\*U(m_Z^2,m_h^2,m_Z^2,m_h^2,m_Z^2)\big]\Big)\notag\\
&\qquad\
-\big[(10+7\*\bar{c}_6)\*m_h^6-2\*(32+19\*\bar{c}_6)\*m_h^4\*m_Z^2+4\*(36+13\*\bar{c}_6)\*m_h^2\*m_Z^4\notag\\
&\qquad\quad+24\*(\bar{c}_6-2)\*m_Z^6\big]\*U(m_Z^2,m_Z^2,m_h^2,m_h^2,m_h^2)\Big]\Big\}\bigg)
\bigg\}~,\\
\label{T_analytic}
  T ={}& \frac{3\*\alpha_e\*\bar{c}_6\*m_h^2}{512\*\pi^2\*s^4\*m_W^4\*(m_h^2 -
    m_W^2)^2\*(m_h^2 - m_Z^2)^2}\*\bigg\{\notag\\
&\ +A(m_h^2)\*\Big((22+9\*\bar{c}_6)\*m_h^2\*(m_h^2-m_W^2)\*m_Z^2\*(m_h^2-m_Z^2)\*s^2\notag\\
&\quad +2\*(m_h^2-2\*m_W^2)\*(m_h^2-m_Z^2)^2\*A(m_W^2)-2\*(m_h^2-m_W^2)^2\*(m_h^2-2\*m_Z^2)\*A(m_Z^2)\notag\\
&\quad
-9\*(2+\bar{c}_6)\*m_h^2\*m_Z^2\*\big[m_h^2\*(m_W^2+m_Z^2)-2\*m_W^2\*m_Z^2\big]\*s^2\*B(m_h^2,m_h^2,m_h^2)\Big)\notag\\
&\ - m_Z^2\*s^2\*\Big(\big[m_h^2\*(m_W^2+m_Z^2)-2\*m_W^2\*m_Z^2\big]\*A(m_h^2)^2\notag\\
&\quad-3\*m_h^2\*\big[2\*(2+\bar{c}_6)\*m_h^4-2\*(1+2\*\bar{c}_6)\*m_W^2\*m_Z^2+(\bar{c}_6-1)\*m_h^2\*(m_W^2+m_Z^2)\big]\*I(m_h^2,m_h^2,m_h^2)\notag\\
&\quad+(m_h^2-m_W^2)\*(m_h^2-m_Z^2)\*\big[(20+9\*\bar{c}_6)\*m_h^4+2\*m_W^2\*m_Z^2\notag\\
&\qquad-2\*m_h^2\*(m_W^2+m_Z^2)-9\*(2+\bar{c}_6)\*m_h^4\*B(m_h^2,m_h^2,m_h^2)\big]\Big)\notag\\
&\ + (m_h^2-m_Z^2)^2\*\Big(-(m_h^2-2\*m_W^2)\*A(m_W^2)^2\notag\\
&\quad+m_W^2\*\Big[A(m_W^2)\*\big[4\*(m_h^2-m_W^2)-9\*(2+\bar{c}_6)\*m_h^2\*B(m_h^2,m_h^2,m_h^2)\big]\notag\\
&\qquad+\big[(20+9\*\bar{c}_6)\*m_h^2-20\*m_W^2\big]\*I(m_h^2,m_h^2,m_W^2)\Big]\notag\\
&\quad+(m_h^4-4\*m_h^2\*m_W^2+12\*m_W^4)\*I(m_h^2,m_W^2,m_W^2)\Big)\notag\\
&\ + (m_h^2-m_W^2)^2\*\Big((m_h^2-2\*m_Z^2)\*A(m_Z^2)^2-(m_h^4-4\*m_h^2\*m_Z^2+12\*m_Z^4)\*I(m_h^2,m_Z^2,m_Z^2)\notag\\
&\quad+m_Z^2\*\Big[A(m_Z^2)\*\big[-4\*m_h^2+4\*m_Z^2+9\*(2+\bar{c}_6)\*m_h^2\*B(m_h^2,m_h^2,m_h^2)\big]\notag\\
&\qquad-\big[(20+9\*\bar{c}_6)\*m_h^2-20\*m_Z^2\big]\*I(m_h^2,m_h^2,m_Z^2)\Big]\Big)\bigg\}~.
\end{align}

\bibliographystyle{JHEP}
\bibliography{references}

\end{document}